\documentclass[12pt]{article}
\usepackage[dvips]{color}
\usepackage{amsmath}
\usepackage{graphicx}
\usepackage{subfigure}
\usepackage{epsfig}
\usepackage{amsmath}
\usepackage{cite}
\usepackage{color}
\usepackage{subfigure}
\def\Box{\hbox{$\rlap{$\sqcup$}\sqcap$}}
\def\Box{\hbox{$\rlap{$\sqcup$}\sqcap$}}
\begin{document}
\begin{center}
\Large{\bf  Swampland dS conjecture in Mimetic $f(R, T)$ gravity}\\
\small \vspace{1cm}{\bf S. Noori Gashti$^{a}$\footnote {Email:~~~saeed.noorigashti@stu.umz.ac.ir}}, \quad
{\bf J. Sadeghi$^{a}$\footnote {Email:~~~pouriya@ipm.ir}}, \quad
{\bf S. Upadhyay$^{b,c,d,e}$\footnote {Email:~~~sudhakerupadhyay@gmail.com}}, \quad
{\bf M. R. Alipour$^{a}$\footnote {Email:~~~mr.alipour@stu.umz.ac.ir}}, \quad
\\
\vspace{0.5cm}$^{a}${Department of Physics, Faculty of Basic
Sciences,\\
University of Mazandaran
P. O. Box 47416-95447, Babolsar, Iran}\\
\vspace{0.2cm}$^{b}${Department of Physics, K.L.S. College,  Nawada, Bihar  805110, India.}\\
\vspace{0.2cm}$^{c}${Department of Physics, Magadh University, Bodh Gaya, Bihar 824234, India}\\
\vspace{0.2cm}$^{d}${Inter-University Centre for Astronomy and Astrophysics (IUCAA),\\ Pune, Maharashtra 411007, India.}\\
\vspace{0.2cm}$^{e}${School of Physics, Damghan University, Damghan  3671641167, Iran.}\\
\small \vspace{1cm}
\end{center}
\begin{abstract}
In this paper, we study a theory of gravity called mimetic $f(R, T)$ in the presence of swampland dS conjecture. For this purpose, we introduce several inflation solutions of the Hubble parameter H(N) from $f(R, T)= R+\delta T$ gravity model, in which R is Ricci scalar, and T denotes the trace of the energy-momentum tensor.
Also, $\delta$ and $N$ are the free parameter and a number of e-fold, respectively. Then we calculate quantities such as potential, Lagrange multiplier, slow-roll, and some cosmological parameters such as $n_{s}$ and $r$. Then we challenge the mentioned inflationary model from the swampland dS conjecture. We discuss the stability of the model and investigate the compatibility or incompatibility of this inflationary scenario with the latest Planck observable data.
\end{abstract}
\newpage
\tableofcontents
\section{Introduction}
The inflationary scenario was proposed to study the expansion of the universe. It expresses how the formation of large scale structures and tries to answer problems such as fine-tuning, horizon, and flatness
\cite{1,2,3} concerning different conditions and conjectures such as slow-roll, constant-roll, etc.
Large-scale and coherent structures of the universe and CMB anisotropies are explained in such a way that the exponential growth of pre-inflationary density fluctuations over a brief period, i.e., approximately $(10^{-33} s)$ can lead to the possibility of creating this structures\cite {4,5,6,7}.
It has recently been proven that a scalar field called $"Inflaton"$ is responsible for inflation\cite{1}. Of course, many potentials for inflation have been studied consistently with observable data. Their cosmological applications have been discussed in detail\cite{8,9,10,11,12}. Modified theories of gravity have always been of interest to researchers in inflation because they do not need to consider the dark matter and dark energy in the face of several cosmological issues such as late-time cosmic acceleration and have been very useful\cite{13,14}. We also remark that in addition to the scalar field in the cosmological inflation scenario, other types of modified gravitational theories have been examined that are consistent with observable data and have discussed various kinds of cosmological consequences such as f(R), f(R, T ), f(G )gravity, also in braneworld models and so on\cite{15,16,17,18,19,20}. We noted that mimetic gravity is an organized Weyl symmetric augmentation of Einstein gravity connected by a singular disformal transformation and explains the evolution of the universe, so that Integrated study from primordial inflation to the late-time acceleration. This mimetic gravity has been investigated in the inflation and solutions of black holes, late-time acceleration, cosmology, and the rotation curves of wormholes, so for further study, you can see in Ref.s.\cite{21,22,23,24,25,26,27,28}. Considering the above concepts, the main purpose of this article is to investigate the solutions of cosmic inflation in the framework of mimetic f (R, T ) gravity with Lagrange multiplier and the potential for the monitoring of swampland conjectures. We first introduce some inflation solutions of the Hubble parameter H(N) and calculate quantities such as potential and Lagrange multiplier and two important cosmological parameters such as $n_{s}$ and $r$, which are called the scalar spectral index and tensor to scalar ratio, respectively. Then by considering one inflation solution of the Hubble parameter, the compatibility or incompatibility of this model of the latest observable data concerning swampland conjectures and analyze the results. This study's exciting and unique point is the combination of mimetic f (R, T) gravity with swampland conjectures in inflation that will have interesting cosmological results and implications. In general,  f(R, T) gravity is introduced in \cite{29} has had many successes in various cosmological studies such as cosmology, Big bang nucleosynthesis,  temporally varying physical constants, density perturbations, viscous cosmology, inflation, late-time acceleration, baryogenesis, and redshift drift. \cite{30,31,32,33,34,35,36,37,38}. Researchers recently studied the swampland conjectures. The swampland program and the weak gravity conjecture have been studied for many cosmological concepts such as inflation, black hole physics, energy and dark matter, brane structure, and other cosmology concepts. These conjectures have been used with other concepts and cosmological conditions, such as various effective theories. The cosmological applications of these studies are well articulated. In the swampland program, we are faced with concepts such as swampland (a set of theories incompatible with quantum gravity) and landscape (a set of theories compatible with quantum gravity). This article also considers the refined swampland conjecture, which we will introduce in detail in this article's continuation. You can see Refe.s.\cite{39,40,41,42,43,44,45,46,47,48,49,50,51,52,53,54,55,56,57,58,59,60,61,62,63,64}. This paper is organized as follows.\\
In Sec 2, we briefly explain the mimetic $f(R, T)$ gravity. In Sec 3, we overview mimetic $f(R, T)$ gravity in an inflationary scenario. In Sec 4, we study several inflation solutions of the Hubble parameters such as $H(N)$ are $(XN^{a}+YN\exp(bN))^{d}$, $(Xc^{N}+YN^{a})^{d}$ and  $(X+YN)^{d}$ from $f(R, T)$ gravity and calculate quantities such as potential, Lagrange multiplier, slow-roll, and other cosmological parameters. In Sec 5, we consider one of the inflationary solutions of the Hubble parameter and study the mimetic $f(R, T)$ inflationary model from the swampland conjecture point of view.
Also, we employ some exciting figures and investigate the compatibility or incompatibility of the inflationary scenario with the latest observable data, i.e.,  BICEP2/Keck Array data, Planck data, and further refining the swampland dS conjecture. In Sec 6, we discuss the stability of the model and finally, we summarize the results with future remarks in Sec 7.

\section{Mimetic $f(R, T)$ Gravity}
This section will explain the mimetic $f(R, T)$ gravity. This section's main purpose is to examine inflation from the point of view of mimetic $f(R, T)$ gravity. By parameterizing the metric using the new degrees of freedom, a wide class of solutions can be created\cite{65,66,67,68,69,70,71,72}. one can introduce the metric$g_{ij}$ as an auxiliary metric $\widetilde{g}_{ij}$ with an auxiliary scalar field that is expressed in the following form, as $g_{ij}=-\widetilde{g}_{ij}\partial_{\alpha}\partial_{\beta}\phi\widetilde{g}^{\alpha\beta}$. We can also give an explanation with respect to the relation expressed as follows, $-(\widetilde{g}_{ij},\phi)\partial_{i}\phi\partial_{j}\phi g^{ij}=1$. With respect to this inflationary model, flat FRW space, and according to the mentioned above and a Lagrange multiplier with a mimetic potential, this model's action is expressed in the following form\cite{72}.

\begin{equation}\label{1}
S=\int\sqrt{-g}d^{4}x\bigg(\mathcal{L}_{m}-V(\phi)+\lambda(g_{ij}\partial_{i}\phi\partial_{j}\phi+1)+f\big(R(g^{ij},T)\big)\bigg),
\end{equation}

\hspace{-0.6cm}where $R$, $T$, $\mathcal{L}_{m}$ are  Ricci scalar, the trace of the energy-momentum tensor and
matter Lagrangian, respectively. The field equation with respect to equation (1) and the  metric $g_{ij}$ which is given by,
\begin{equation}\label{2}
\begin{split}
&\frac{1}{2}g_{ij}(\lambda(g^{\alpha\beta}\partial_{\alpha}\phi\partial_{\beta}\phi+1)-V(\phi))+\frac{1}{2}g_{ij}f(R, T)+\nabla_{i}\nabla_{j}f_{R}-R_{ij}f_{R}\\
&-\lambda\partial_{i}\phi\partial_{j}\phi+\frac{1}{2}T_{ij}-f_{T}(T_{ij}+\Pi_{ij})-g_{ij}\Box f_{R}=0,
\end{split}
\end{equation}

\hspace{-0.6cm}where $\Pi_{ij}=g^{\alpha\beta}\frac{\delta T_{\alpha\beta}}{\delta g^{ij}}$ and $\Box=\nabla^{i}\nabla_{i}$, $\nabla_{i}$, are the d’Alembertian operator and covariant derivative, also $f_{R}$ and $f_{T}$ denote partial derivatives of $f(R, T)$ with respect to $R$ and $T$, respectively\cite{72}. The energy-momentum tensor for the cosmos filled with perfect fluid is expressed in the following form.

\begin{equation}\label{3}
\begin{split}
T_{ij}=p_{m}g_{ij}+(p_{m}+\rho_{m})v_{i}v_{j},
\end{split}
\end{equation}

\hspace{-0.6cm}where $p_{m}$, $\rho_{m}$ and $v_{i}$ are the pressure and energy density and the four velocity, respectively. Also the scalar curvature $R$ can be expressed as $R=6(2H^{2}+\dot{H})$ and $T=\rho-3p$, If assumed $\mathcal{L}_{m}=-p_{m}$, one can obtain\cite{72},

\begin{equation}\label{4}
\begin{split}
&\frac{1}{2}g_{ij}\big(\lambda(g^{\alpha\beta}\partial_{\alpha}\phi\partial_{\beta}\phi+1)-V(\phi)\big)+\frac{1}{2}g_{ij}f(R, T)+\nabla_{i}\nabla_{j}f_{R}-R_{ij}f_{R}\\
&-\lambda\partial_{i}\phi\partial_{j}\phi+\frac{1}{2}T_{ij}-f_{T}(T_{ij}+p_{m}g_{ij})-g_{ij}\Box f_{R}=0.
\end{split}
\end{equation}

\hspace{-0.6cm}With respect to equations (1), (2) and the auxiliary scalar field $\phi$, we will have,

\begin{equation}\label{5}
\begin{split}
-V'(\phi)-2\nabla^{i}(\lambda\partial_{i}\phi)=0,
\end{split}
\end{equation}

\hspace{-0.6cm}prime represents the derivative concerning $\phi$, and with respect to equations (1), (2), and the Lagrange multiplier $\lambda$, we arrive,
\begin{equation}\label{6}
\begin{split}
g^{ij}\partial_{i}\phi\partial_{j}\phi=-1.
\end{split}
\end{equation}

\hspace{-0.6cm}The Friedmann equations for the FRW background with the presumption that the scalar field depends on t is given by the following equation\cite{72}.

\begin{equation}\label{7}
\begin{split}
6(\dot{H}+H^{2})f_{R}-f(R,T)-6H\frac{\textrm{d}f_{R}}{\textrm{d}t}-\lambda(\dot{\phi}^{2}+1)-2f_{T}p_{m}+V(\phi)+\rho_{m}(2f_{T}+1)=0,
\end{split}
\end{equation}

\begin{equation}\label{8}
\begin{split}
f(R, T)-2(\dot{H}+3H^{2})f_{R}+p_{m}(4f_{T}+1+4H\frac{\textrm{d}f_{R}}{\textrm{d}t})-\lambda(\dot{\phi}^{2}+1)+2\frac{\textrm{d}^{2}f_{R}}{\textrm{d}t^{2}}-V(\phi)=0,
\end{split}
\end{equation}

\begin{equation}\label{9}
\begin{split}
6H\lambda\dot{\phi}-V'(\phi)+2\frac{\textrm{d}(\lambda\dot{\phi})}{\textrm{d}t}=0,
\end{split}
\end{equation}

\begin{equation}\label{10}
\begin{split}
\dot{\phi}^{2}-1=0,
\end{split}
\end{equation}

\hspace{-0.6cm}where the dot represents the time derivative. By assuming the $\phi=\phi(t)$ the equation (7) convert to,

\begin{equation}\label{11}
\begin{split}
-2(\dot{H}+3H^{2})f_{R}+f(R, T)+4H\frac{\textrm{d}f_{R}}{\textrm{d}t}+2\frac{\textrm{d}^{2}f_{R}}{\textrm{d}t^{2}}-V(t)+p_{m}(4f_{T}+1)=0.
\end{split}
\end{equation}

\hspace{-0.6cm}So the mimetic potential in $f(R, T)$ which is given by

\begin{equation}\label{12}
\begin{split}
V(t)=f(R, T)+4H\frac{\textrm{d}f_{R}}{\textrm{d}t}+2\frac{\textrm{d}^{2}f_{R}}{\textrm{d}t^{2}}+p_{m}(4f_{T}+1)-2(\dot{H}+3H^{2})f_{R}.
\end{split}
\end{equation}

\hspace{-0.6cm}According to the above concepts, by selecting appropriate models from $f(R, T )$ gravity,  one can reconstruct the mimetic potential in equation (12). We can also solve equation (7) with respect to the Lagrange multiplier that we will have.
\begin{equation}\label{13}
\begin{split}
\lambda(t)=3(\dot{H}+H^{2})f_{R}-\frac{1}{2}f(R, T)-3H\frac{\textrm{d}f_{R}}{\textrm{d}t}+\rho_{m}(f_{T}+\frac{1}{2})+\frac{1}{2}V(t)-f_{T}p_{m}.
\end{split}
\end{equation}

\section{Inflation Setup}

After an overview of mimetic $f(R, T)$ gravity, now in this section, we want to study the effect of several different inflation models within the mimetic $f(R, T )$ gravity structure and compare it with the latest Planck observable data\cite{74, J}. Hence we choose the simplest form of the $f(R, T)$ gravity model, i.e., $f(R, T )=R+\delta T$. Then we introduce $n_{s}$, $r$ and slow-roll parameters in this structure\cite{72}. these parameters can be easily expressed in terms of the number of e-folds using the following equations, so

\begin{equation}\label{14}
\begin{split}
\frac{\textrm{d}}{\textrm{d}t}=H(N)\frac{\textrm{d}}{\textrm{d}N},\hspace{1cm}\frac{\textrm{d}^{2}}{\textrm{d}t^{2}}=H(N)\frac{\textrm{d}H}{\textrm{d}N}+H^{2}(N)\frac{\textrm{d}^{2}}{\textrm{d}N^{2}}.
\end{split}
\end{equation}

\hspace{-0.6cm}So the slow-roll parameters in terms of $N$, which is given by,
\begin{equation}\label{15}
\begin{split}
\epsilon=-\frac{H(N)}{4H'(N)}(\frac{6\frac{H'(N)}{H(N)}+\frac{H''(N)}{H(N)}+(\frac{H'(N)}{H(N)})^{2}}{\frac{H'(N)}{H(N)}+3})^{2},
\end{split}
\end{equation}

\begin{equation}\label{16}
\begin{split}
\eta=-(\frac{\frac{1}{2}(\frac{H'(N)}{H(N)})^{2}+9\frac{H'(N)}{H(N)}+3\frac{H''(N)}{H(N)}+3\frac{H''(N)}{H'(N)}-\frac{1}{2}(\frac{H''(N)}{H'(N)})^{2}+\frac{H''(N)}{H'(N)}}{2(\frac{H'(N)}{H(N)}+3)}),
\end{split}
\end{equation}

\hspace{-0.6cm}where primes represent a derivative of N. The scalar spectral index and tensor-to-scalar ratio which is given by,

\begin{equation}\label{17}
\begin{split}
n_{s}=1+2\eta-6\epsilon, \hspace{1cm}r=16\epsilon.
\end{split}
\end{equation}

\hspace{-0.6cm}One can consider the $n_{s}$ and $r$ restrictions resulted from the marginalized joint 68\% and 95\% CL regions of the Planck 2018 in compound with BK14+BAO data, viz $n_{s}=0.9649\pm0.0042$ at 68\% Cl, and $r<0.1$ at 95\% CL and from BICEP2/Keck Array BK14 recent data $r<0.056$ at 95\% CL \cite{J}. The exact value of the mimetic potential and Lagrange multiplier can also be determined according to equation (14) and employ the mimetic $f(R, T )$ gravity. In the following, we introduce several inflation solutions of the Hubble parameter H(N) and use the method expressed in these two sections. We obtain the exact values for each cosmological parameter,  mimetic potential, and Lagrange multiplier. Then,  we examine these results from the swampland conjecture perspective and compare them with observable data.
\section{Several inflation solutions of Hubble parameters}
This section considers three impressive solutions of Hubble parameters named models I, II, and III, respectively.
\subsection{Model I}
For the first inflation model, the Hubble rate is given by the following function of the number of e-folds:

\begin{equation}\label{18}
\begin{split}
H(N)=(Xc^{N}+YN^{a})^{d},
\end{split}
\end{equation}

\hspace{-0.6cm}where $X, Y, c, a$ and $ d$ are free parameters and $N$ is the number of e-folding. the $\epsilon$, $\eta$, $n_{s}$ and $r$ are calculated as shown in Appendix A (Eq.s(53-56)).

By determining each of these free parameters' values, we confirm the correspondence of each of these inflation scenarios with the observable data. According to the mentioned point, we consider the values of each of these free parameters in this form.
\begin{equation*}\label{19}
 Y=50,\ X=-1,\ c=1,\ a=-0.0005,\ d=1.
\end{equation*}

\hspace{-0.6cm}Hence,  the two cosmological parameters, i.e., the scalar spectral index and the tensor-to-scalar ratio for these free parameters,

\begin{equation*}\label{20}
r=0.000152005, \hspace{1cm} n_{s}=0.966221,
\end{equation*}

\hspace{-0.6cm}which correspond to the observable data \cite{74}. It can be noted that after examining each of these cosmological parameters and comparing them with observable data, the values of mimetic potential and the Lagrange multiplier responsible for producing the inflationary model (18) can also be given according to the equations (18), (12), (13) and (14) in mimetic f(R, T) gravity. So we will have

\begin{equation}\label{19}
\begin{split}
V=&-\frac{(-2+3(-1+\delta)\delta)d(YN^{a}+Xc^{N})^{-1+d}(YN^{a}a+XNc^{N}\log(c))(YN^{a}+Xc^{N})^{d}}{N(1+\delta)(1+3\delta)}\\
&+\frac{3(1+\delta)(YN^{a}+Xc^{N})^{2d}}{1+3\delta}-\delta(-3+\rho),
\end{split}
\end{equation}

\begin{equation}\label{20}
\begin{split}
\lambda =&\frac{1}{2}\bigg(-\frac{3d\delta^{2}(YN^{a}+Xc^{N})^{-1+d}(YN^{a}a+XNc^{N}\log(c))(YN^{a}+Xc^{N})^{d}}{N(1+4\delta+3\delta^{2})}\\
&-\frac{3(2+5\delta)(YN^{a}+Xc^{N})^{2d}}{1+3\delta}-3+\delta+(2+\delta)\rho\bigg).
\end{split}
\end{equation}

\subsection{Model II}
This part follows a similar process to the last detail; hence we consider another inflationary model in which the Hubble rate describes the cosmological evolution.
\begin{equation}\label{21}
\begin{split}
H(N)=(XN^{a}+YN\textrm{e}^{b N})^{d},
\end{split}
\end{equation}

\hspace{-0.6cm}where $X, Y, b, a$ and $ d$ are free parameters. The slow-roll and spectral parameters of model II are obtained as shown in Appendix B (Eq.s(57-60)). Now, we consider the free parameters for the second inflation model as

\begin{equation*}\label{23}
Y=25,\ X=-1.5,\ b=-0.025,\ a=-0.09,\ d=0.1.
\end{equation*}
Hence, the scalar spectral index and the tensor-to-scalar ratio obtain

\begin{equation*}\label{23}
n_{s}=0.965577, \hspace{1cm}r=0.0155542,
\end{equation*}

\hspace{-0.6cm}which correspond to the observable data \cite{74}. like model I, the values of mimetic potential and the Lagrange multiplier are calculated as

\begin{equation}\label{22}
\begin{split}
V=&-\frac{1}{N(1+\delta)(1+3\delta)}\bigg(XN^{a}(YN\textrm{e}^{Nb}+XN^{a})^{-1+2d}\bigg[9N(-1+\delta)(1+2\delta)(1+\delta)\\
&+a\Big(-4+\delta\big(-5+\delta(11+6\delta)\big)\Big)d\bigg]+YN\textrm{e}^{b N}(Y\textrm{e}^{b N}+XN^{a})^{-1+2d}\bigg(9N(-1+\delta)(1+\delta)\\
&\times (1+2\delta)+(1+Nb)\Big(-4+\delta\big(-5+\delta(11+6\delta)\big)\Big)d\bigg)+N\delta(1+\delta)(1+3\delta)\rho\bigg),
\end{split}
\end{equation}

\begin{equation}\label{23}
\begin{split}
\lambda=&-3Y^{2}N^{2}\textrm{e}^{2Nb}(-1+\delta)-3X^{2}N^{2a}(-1+\delta)-6AY\textrm{e}^{Nb}N^{1+a}(-1+\delta)\\
&-{(YNe^{b N}+XN^{a})^{d}\Big(Xa N^{a -1}+Y \textrm{e}^{Nb}(1+Nb)\Big)(-3+\delta)d}\\
&-\frac{1}{2N(1+\delta)(1+3\delta)}(Y\textrm{e}^{Nb}N+XN^{a})^{-1+2d}(-2+5\delta)\\
&\times\bigg[XN^{a}(3N(1+\delta)+a(2+3\delta)d)+YN\textrm{e}^{Nb}\Big(3N(1+\delta) +(1+Nb)(2+3\delta)d\Big)\bigg]\\
&+\rho+\frac{\delta \rho}{2}.
\end{split}
\end{equation}

\subsection{Model III}

In the final inflation model, we consider a subclass of the cases studied above, where the following Hubble parameter describes the cosmological evolution. In this model, $X$, $Y$, and $d$ are constants parameters. the important point, i.e., the constants parameter are $Y<0$, $X > 0$. So, we have $|\dot{H}|\ll H^{2} $ during the inflationary duration, which causes $\frac{X}{Y}\gg N$, such that the Hubble parameter is roughly constant (de Sitter) during the inflation. After inflation finishes, the second sentence evolves important, and the Hubble rate decays.

\begin{equation}\label{24}
\begin{split}
H(N)=(X+YN)^{d}.
\end{split}
\end{equation}

\hspace{-0.6cm}We can investigate The slow-roll parameter in equations (15) and (16) in the following form.

\begin{equation}\label{25}
\begin{split}
\epsilon=-\frac{Yd(6X+Y(-1+6N+2d))^{2}}{4(X+YN)(3X+Y(3N+d))^{2}},
\end{split}
\end{equation}

\begin{equation}\label{26}
\begin{split}
\eta=\frac{Y\Big(Y+8YN+X(8-26d)-2Yd(-2+13N+3d)\Big)}{4(X+NY)(3X+Y(3N+d))}.
\end{split}
\end{equation}

\hspace{-0.6cm}The scalar spectral index and the tensor-to-scalar ratio in equations (17) are obtained as,

\begin{equation}\label{27}
\begin{split}
n_{s}=&1+\frac{3Yd(6X+Y(-1+6N+2d))^{2}}{2(X+YN)(3X+Y(3N+d))^{2}}\\
&+\frac{Y\Big(Y+8YN+X(8-26d)-2Yd(-2+13N+3d)\Big)}{2(X+YN)(3X+Y(3N+d))},
\end{split}
\end{equation}

\begin{equation}\label{28}
\begin{split}
r=-16\frac{Yd(6X+Y(-1+6N+2d))^{2}}{4(X+YN)(3X+Y(3N+d))^{2}}.
\end{split}
\end{equation}

\hspace{-0.6cm}Like the previous part, we determine these free parameters, i.e.,

\begin{equation*}\label{23}
X=0.75,\ Y=-0.00699999,\ d=0.2.
\end{equation*}

\hspace{-0.6cm}So, the scalar spectral index and the tensor-to-scalar ratio obtain respect to these free parameters.

\begin{equation*}\label{23}
n_{s}=0.964569, \hspace{1cm}r=0.06836,
\end{equation*}

\hspace{-0.6cm}where correspond to the observable data \cite{74}. Also, the mimetic potential and the Lagrange multiplier are calculated as

\begin{equation}\label{29}
\begin{split}
V=&\frac{-9(X+YN)^{-1+2d} (X+YN)(-1+\delta)(1+2\delta)(1+\delta)+Y\Big(4+\delta(5-\delta(11+6\delta))\Big)d }{(1+\delta)(1+3\delta)}-\delta\rho,
\end{split}
\end{equation}

\begin{equation}\label{30}
\begin{split}
\lambda=& \frac{(X+YN)^{-1+2d}(-3(X+YN)(1+\delta)\Big(-4+\delta+6\delta^{2}\Big)+Y\Big(10+\delta\big(18-\delta(5+6\delta)\big)\Big)d}{2(1+\delta)(1+3\delta)}\\
&+(2+\delta)\rho.
\end{split}
\end{equation}

\section{ Mimetic $f(R, T)$ in RSC Perspective }

In this section, we examine one of the inflation solutions of the Hubble parameter H(N),i.e., model (iii) from the swampland conjecture point of view and compare their results with observable data\cite{74}. Researchers have recently studied swampland conjectures as an exciting subject for examining cosmological structures such as black holes, dark matter, dark energy, inflation, etc. Also their cosmological implications have been thoroughly reviewed \cite{55,56,57,58,59,60,61,62,63,64}.
But a significant point to note about swampland conjectures is that it still struggles with many problems, For example, the discrepancy between single-field slow-roll inflation and swampland dS conjecture.\cite{60}
Of course, solutions have always been proposed to solve the existing problems and contradictions, such as considering a Gauss-Bonnet term to calculate the slow-roll parameters, which leads to resolving the mentioned contradiction\cite{60}.
Therefore, they are always looking for a deeper study and matching the mentioned conjectures with different concepts of cosmology with other approaches.
One of these techniques is, in fact, the correction of conjectures themselves. For example, a new conjecture called further refined swampland dS conjecture has recently been introduced, which has been able to answer many contradictions in various studies of cosmology\cite{a,b}.
Of course, swampland conjectures, in many cases, are in good agreement with the concepts, and the latest observable data, including the warm inflation\cite{c}.
Also, recently, many efforts have been made to generalize the swampland program. Many conjectures have been added to it, including gravitino swampland conjecture\cite{d}.
Of course, one of the most important motivations in the study of different concepts of cosmology according to the swampland conjectures can be pointed out in the inflation, for example, by examining different types of inflation models in various frameworks and also other types of modified theories of gravity such as f(R), f(R, T), f(G), etc., according to the structure of swampland conjectures and analysis of results and their comparison with the latest observable data can achieve a new classification of inflation models.
 It is even possible to determine the best model from many models through these adaptations.
But despite all the above explanations, this statement is still incomplete, and as we mentioned, it still faces serious problems and challenges\cite{40,41,42,43,44,45,46,47,48,49}.
Therefore, this article intends to challenge a kind of inflation model with recent conjectures.
Of course, the swampland program, which has recently emerged from string theory, has been introduced to answer a long-standing problem, quantum gravity.
Among the conjectures used in this program, we can mention distance conjecture, dS conjecture, weak gravity conjecture, and TCC\cite{50,51,52,53,54,55,56,57,58,59,60}.
In this statement, we also use concepts such as the landscape, that is, a set of theories consistent with quantum gravity, and a broader area surrounding the landscape, which expresses a set of approaches that are not compatible with quantum gravity and considered as swampland.
Therefore,  one introduced a series of mathematical formulations to present these conjectures, such as the dS swampland conjecture. The satisfaction of these equations shows the compatibility of the models with the introduced conjectures.
In other words, the swampland program expresses a series of conventions and criteria that the adherence of different theories to these criteria leads to their compatibility with quantum gravity.
These concepts provide very powerful tools for examining the compatibility of different cosmological concepts and structures with the ideas mentioned and the latest observable data\cite{e}
The dS swampland conjecture mentioned in the following equation is always used in many cosmic concepts, different types of phenomenological models, string theory, etc., by expressing the constraint on potential\cite{50,51,52,53,54,55,56,57,58,59,60}.
Also, these conjectures about different types of the most important cosmology parameters, such as scalar spectrum index and tensor to scalar ratio, which are among the most important cosmological parameters in studying different inflation models, have had interesting results. In this article, we intend to challenge them with new restrictions.
One of the most important motivations in these studies could be to find models consistent with these conjectures, which in some way leads to compatibility with quantum gravity, which may be used to consider new solutions to study quantum gravity.
Of course, these arguments are theoretical, profoundly consistent with some concepts, and related to the latest observable data. This article considers the refined swampland conjecture, which is expressed in the following form.

\begin{equation}\label{31}
|\nabla V|\geq\frac{C_{1}}{M_{p}}V, \hspace{12pt}  \frac{min \nabla\partial V\leq}{V} -\frac{C_{2}}{M_{pl}^{2}}V,
\end{equation}

The above equations for the $V>0$, which is given by

\begin{equation*}\label{57}
\sqrt{2\epsilon_{V}}\geq C_{1} ,\hspace{12pt}  or \hspace{12pt} \eta_{V}\leq -C_{2},
\end{equation*}

\hspace{-0.6cm}where $C_{1}$ and $C_{2} $ are both positive constants of the order of $\mathcal{O}(1)$.
We consider the mimetic potential of inflation solution of the Hubble parameter H(N) in equation (24), and with respect to the mimetic potential (29) and the scalar field as a function of $N$ that is defined as $\frac{\textrm{d}N}{\textrm{d}t}=H(N)$ and $\phi=t$, one can obtain,

\begin{equation}\label{32}
\begin{split}
V(\phi)=&-\delta\rho+\frac{1}{(1+\delta)(1+3\delta)}\bigg(\Big((X+Y)^{1-d}-Y(-1+d)\phi^{\frac{1}{1-d}}\Big)^{-1+2d}\\
&\times\bigg[Y\Big(4+\delta\big(5-\delta(11+6\delta)\big)\Big)d-9(-1+\delta)(1+\delta)(1+2\delta)\\
&\times\Big((X+Y)^{1-d}-Y(-1+d)\phi\Big)^{\frac{1}{1-d}}\bigg]\bigg).
\end{split}
\end{equation}

\begin{figure}[h!]
\begin{center}
\subfigure[]{
\includegraphics[height=6cm,width=6cm]{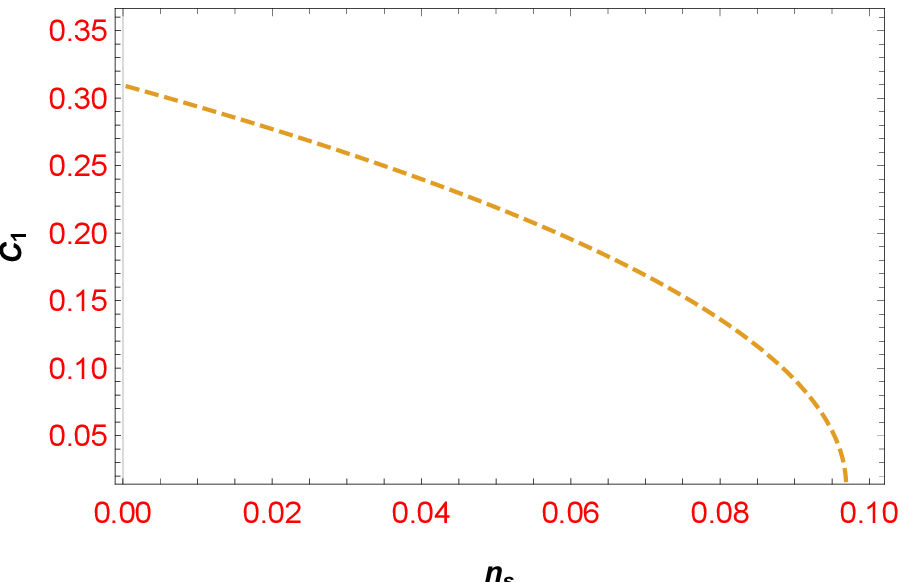}\label{1a}}
\subfigure[]{\includegraphics[height=6cm,width=6cm]{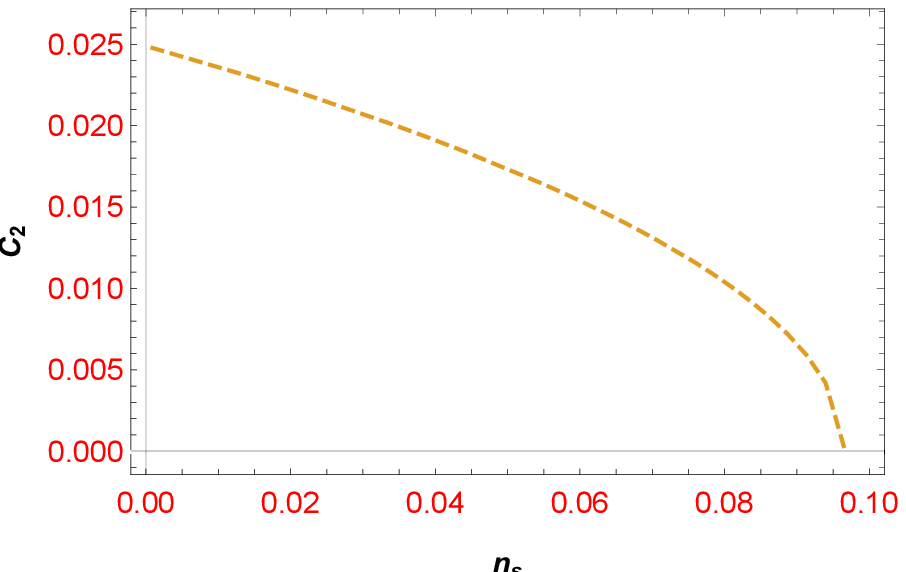}\label{1b}}
\caption{\small{The plot of $C_{1}$ and $C_{2}$ in terms of $n_{s}$ }}\label{1}
\end{center}
\end{figure}

\begin{figure}[h!]
\begin{center}
\subfigure[]{
\includegraphics[height=6cm,width=6cm]{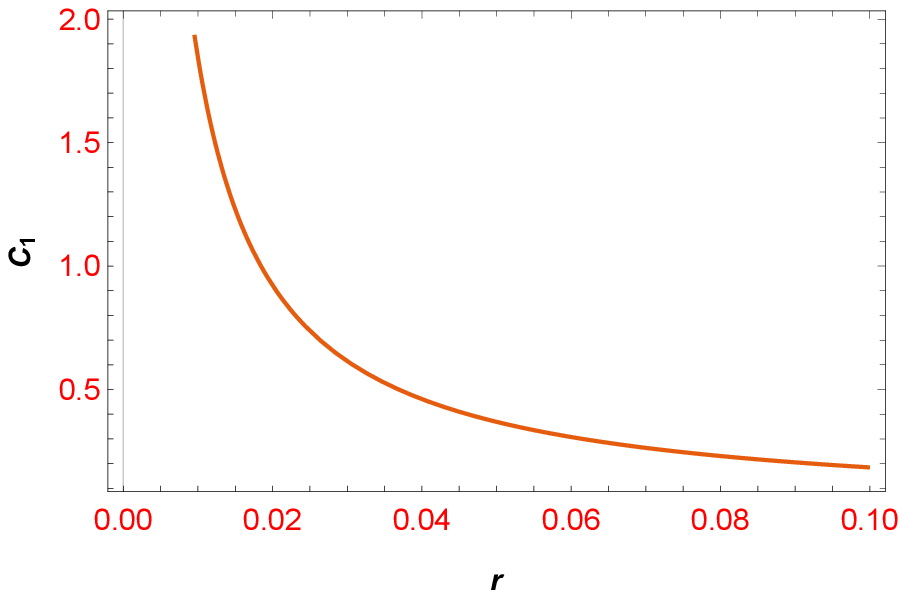}
\label{2a}}
\subfigure[]{
\includegraphics[height=6cm,width=6cm]{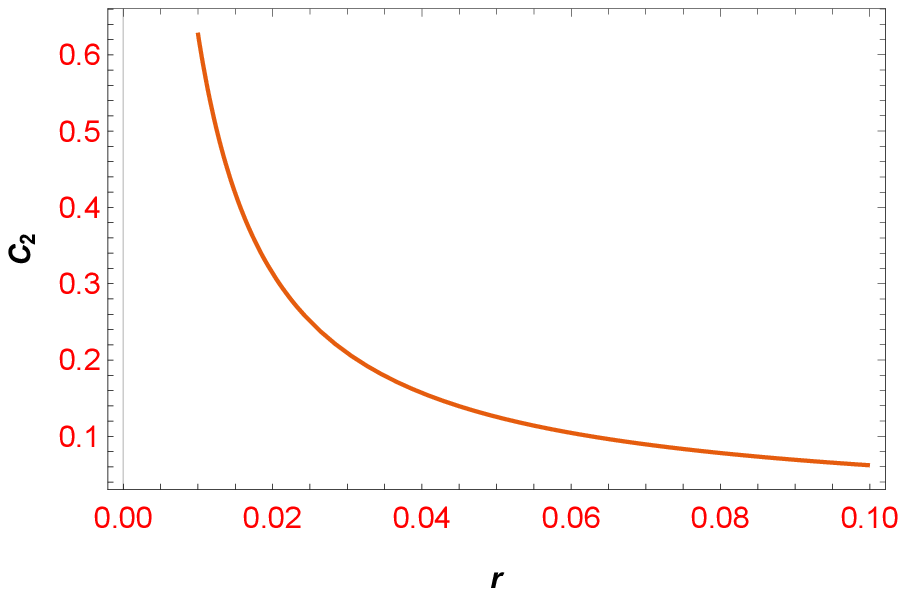}
\label{2b}}
\caption{\small{The plot of $C_{1}$ and $C_{2}$ in terms of $r$ }}
\label{2}
\end{center}
\end{figure}

\hspace{-0.6cm}Other cosmological quantities and parameters can also be reconstructed in this way. According to equation (32), the first and second derivatives of the potential and swampland dS conjecture are calculated as shown in Appendix C (Eq.s(61-64)). We use a specific process to examine a series of constraints and compare them to observable data \cite{74}. Using the condition as $\frac{\textrm{d}N}{\textrm{d}t}=H(N)$ and $\phi=t$, we rewrite equations (\ref{27}) and (\ref{28}) and then we invert these equations as $(\phi-n_{s})$ and $(\phi-r)$. The descriptions and reconstructed values of these two cosmological parameters are expressed as
\begin{equation}\label{33}
\begin{split}
n_{s}&= 1+\frac{24Yd\Big((X+Y)^{1-d}-Y(-1+d)\phi\Big)^{\frac{3}{-1+d}}\bigg[Y-2Yd-6\Big((X+Y)^{1-d}-Y(-1+d)\phi\Big)^{\frac{1}{1-d}}\bigg]^{2}}{(3+Yd\Big((X+Y)^{1-d}-Y(-1+d)\phi\Big)^{\frac{1}{-1+d}})^{2}}\\
&+\frac{Y\Big((X+Y)^{1-d}-Y(d-1)\phi\Big)^{\frac{1}{-1+d}}\bigg(8-26d
+Y( 1+4d -6d^{2})\Big((X+Y)^{1-d}-Y(d-1)\phi\Big)^{\frac{1}{d-1}}\bigg)}{2\bigg(3+Yd\Big((X+Y)^{1-d}-Y(-1+d)\phi\Big)^{\frac{1}{d-1}}\bigg)},
\end{split}
\end{equation}
and
\begin{equation}\label{34}
r=-\frac{4Yd\Big((X+Y)^{1-d}-Y(-1+d)\phi\Big)^{\frac{3}{-1+d}}\Big(Y-2Yd-6\Big((X+Y)^{1-d}-Y(-1+d)\phi\Big)^{\frac{1}{1-d}}\Big)^{2}}{\Big(3+Yb\Big((X+Y)^{1-d}-Y(-1+d)\phi\Big)^{\frac{1}{-1+d}}\Big)^{2}}.
\end{equation}
Here, we examined one of the inflation solutions of the Hubble parameter $H(N)$ and studied the mimetic $f(R, T)$ inflationary model from the swampland conjectures point of view. We investigated this inflationary scenario's compatibility or incompatibility with the latest observable data, i.e.,  Planck data and BICEP2/Keck Array data. For this purpose, the limitations of the swampland conjecture were determined in terms of two cosmological parameters. $n_{s}$  and $r$ are obtained for all the considered inflation models according to the selected values for free parameters. They were consistent with Planck's observable data \cite{74}). We obtained these two cosmological parameters in terms of $ \phi$. By placing each of these structures  as $(\phi-n_{s})$ and $(\phi-r)$ in equations (\ref{C3}) and (\ref{C4}), new structures are formed, i.e., $ (C_{1}-n_{s}) $, $(C_{2}-n_{s})$, $(C_{1}-r)$ and $(C_{2}-r)$.  Then, we draw some figures for these structures with respect to Planck's observable data and validated values of free parameters for (model III).
The plots of $C_1$ and $C_2$ in terms of $n_s$ are depicted in FIGS. \ref{1a}
and \ref{2a}.

As shown in figure (1), the restriction of swampland conjectures is determined in terms of the scalar spectral index, which used the validated values for the free parameters for this inflation model obtained in the previous section. As you can see, the acceptable values are specified for the scalar spectral index and the constant coefficient of swampland conjectures ($C_{1}$) and ($C_{2}$). In literature, the constants swampland conjecture, ($C_{1}$) and ($C_{2}$) are the positive value of unit order. You can see these issues in the figures. The validated values for the ($C_{1}$) and ($C_{2}$) are shown in the figures. ($C_{2}$) values is less than the ($C_{1}$). Similarly, the swampland conjecture's limitations for this inflation model in terms of the tensor-to-scalar ratio are also shown in figure (2). The validated values for this cosmological parameter and the constant coefficients of the swampland conjecture are well specified. We also benefited from the accepted values of free parameters for this inflation model. Other restrictions can also be considered below. Another type of constraint is the study of changes in two cosmological parameters, i.e., the tensor-to-scalar ratio in terms of the scalar spectral index with respect to mimetic $f(R, T)$ gravity.
\begin{figure}[h!]
 \begin{center}
 \subfigure[]{
 \includegraphics[height=6cm,width=6cm]{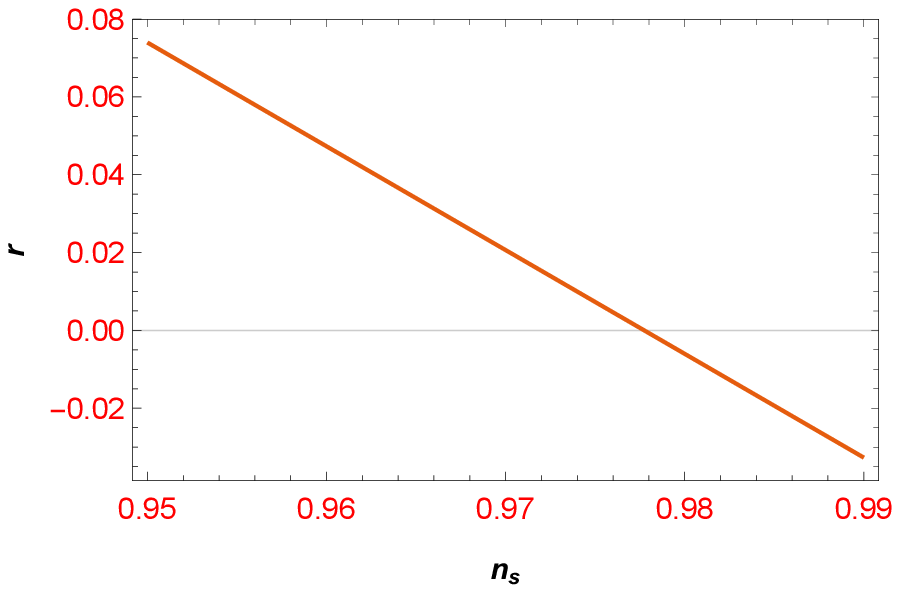}
 \label{3}}
 \caption{The plot of $r$ in term of $n_{s}$}
 \label{3}
 \end{center}
 \end{figure}

Another limitation is shown in figure (3) about the changes in the two cosmological parameters, namely the tensor-to-scalar ratio in terms of the scalar spectral index. The allowable range of both parameters is well specified in this plot. The critical point is to examine these changes according to the validated values for the mentioned inflation model and compare them with observable data. One can discuss all these limitations concerning other mimetic $f(R, T)$ gravity and other models as $f(G)$ gravity and compare the results of these studies. It can lead to exciting developments and even determine which theories are more concrete and more in line with observable data.
Also, to confirm our arguments about the swampland conjecture and the calculations performed, we consider the observable data such as $r$, $n_{s}$,$C_{1}$, $C_{2}$ and the constant parameters that are determined for each model in this paper. We study new restrictions by forming a function as $f\geq C_{1}^{2}C_{2}^{2}$ and determine the allowable range of the swampland component, and specify the $C_{i}s$ that can be acceptable for the model.

 \begin{figure}[h!]
 \begin{center}
 \includegraphics[height=6cm,width=6cm]{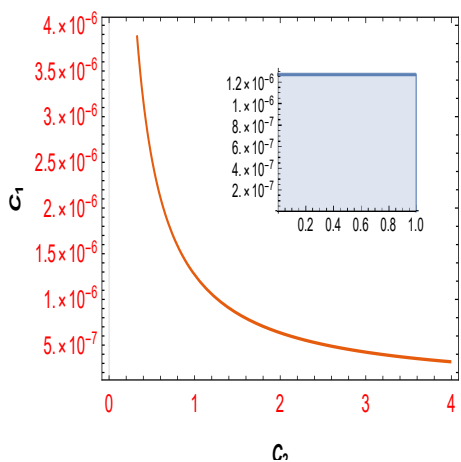}
 \caption{The resulting constraints for $C_{1}$  and $C_{2}$ with respect to $r$, and
$n_{s}$ and constant parameters. According to the calculations performed in the text, the allowable range for this model pick for the vertical axis contains values smaller than $1.26911\times10^{-6}$ in terms of a permissible range for the horizontal axis, i.e., the interval [0, 1].
This allowable area is determined as a small window in the figure4}
 \label{fig4}
 \end{center}
 \end{figure}
In general, as shown in figure \ref{fig4}, the allowable range for the swampland conjecture is determined according to the observable data $r$, $n_{s}$ and the constant parameters. That is, it can be concluded that $C_{i}\leq1.26911\times10^{-6}$(the allowable range for these calculations is in a small window in figure 4). can be the allowable values for this model, which can be another confirmation of the alignment of the model about the swamp conjectures.

\subsection{further refining swampland conjecture}
According to the explanations we provided about recent conjectures and the equations mentioned in the previous section, we can use further refining swampland conjecture conditions to examine any effective low-energy theory of gravity concerning $M_p=1$. \cite{b,100}. A new conjecture that a combination of two relations in equation (31) called further refining swampland conjecture was introduced by Andriot, and Roupec, which is expressed as.\cite{a}.

\begin{equation*}\label{57}
\bigg(M_{p}\frac{\nabla V}{V}\bigg)^{q}-aM_{p}^{2}\frac{min \nabla_{j}\nabla_{j}V}{V}\geq b \hspace{12pt} a+b=1, \hspace{12pt} a,b>0, \hspace{12pt}q>2.
\end{equation*}

\hspace{-0.6cm}The above equations can also be rewritten in terms of slow-roll parameters in the following form.

\begin{equation*}\label{57}
\big(2\epsilon_{V}\big)^{q/2} -a\eta_{V}\geq b.
\end{equation*}

\hspace{-0.6cm}In connection with the recent conjecture, some works have been done in the literature, which you can see for more review in \cite{b,100}. With respect to the above equations, we will have,
 \begin{equation}\label{135}
(F_1)^{\frac{q}{2}}-a F_2 \geq b,
\end{equation}
where the parameters $a,b$ and $q$ are free parameters. Also $F_1$ and $F_2$ are defined as follows
\begin{equation}\label{136}
F_1=\frac{|\textrm{d}V(\phi)/\textrm{d}\phi|}{V(\phi)}=\sqrt{2\epsilon_V} \hspace{1.5cm},   F_2=\frac{\textrm{d}^2V(\phi)/\textrm{d}\phi^2}{V(\phi)}=\eta_V.
\end{equation}
Now, we rewrite $F_1$ and $F_2$ in the slow-roll region according to the parameters $r$ and $n_s$, the tensor-to-scalar ratio, and the spectral index of the primordial curvature power spectrum, respectively.
\begin{equation}\label{137}
F_1=\sqrt{2\epsilon_V}=\sqrt{\frac{r}{8}}, \hspace{1.5cm}   F_2=\eta_V=\frac{1}{2}(n_s-1+\frac{3r}{8}).
\end{equation}
Here, we examine whether these three inflation models within the mimetic f(R; T) gravity structure can satisfy further refining swampland dS conjecture.
\subsubsection{Model I}
For the first model The scalar spectral index and the tensor-to-scalar
ratio, are obtained  $n_{s}=0.966221$ and  $r=0.000152005$. By placing these values in Equation \eqref{137}, we find
\begin{equation}\label{138}
F_1=\sqrt{2\epsilon_V}=\sqrt{\frac{r}{8}}=0.00436,  \hspace{1.5cm} F_2=\eta_V=\frac{1}{2}(n_s-1+\frac{3r}{8})=-0.01686.
\end{equation}
Considering the refined swampland conjecture \eqref{31}, we find
\begin{equation}\label{139}
C_{1}\leq 0.00436,  \hspace{1.5cm}   C_2 \leq 0.01686,
\end{equation}
since $C_1$ and $C_2$ are not unit orders ($C_1=C_2 \neq \mathcal{O}(1)$), they contradict the refined swampland conjecture.
We now examine the first model by further refining the swampland conjecture. By placing the equation \eqref{138} into \eqref{135}, we obtained
\begin{equation}\label{140}
(0.00436)^q+ 0.01686 a \geq 1-a \hspace{1cm}, or \hspace{1cm} (0.00436)^q \geq 1- 1.01686 a.
\end{equation}
We can find $a$ to satisfy the condition
\begin{equation}\label{141}
\frac{1}{1.01686}[1-(0.00436)^q]\leq a <1 ,  \hspace{1cm} q>2,
\end{equation}
therefore, further refining the swampland conjecture can be satisfying. Equation \eqref{140} holds for all values of $q > 2$ with respect to $a < \frac{1}{1.01686}= 0.98342 $. We can give an example for $a,b,q$ variables that met for this model. By selecting $q=2.1$, and using relations \eqref{140} and \eqref{141} we have $0.98340 \leq a < 0.98342 $ and by investigating the $a=0.98341$, we obtain $b=1-a=0.01659$.
\subsubsection{Model II}
For the second model, the scalar spectral index and the tensor-to-scalar ratio, are obtained  $n_{s}=0.965577$ and $r=0.0155542$. By placing these values in equation \eqref{137}, we find
\begin{equation}\label{142}
F_1=\sqrt{2\epsilon_V}=\sqrt{\frac{r}{8}}=0.044094,  \hspace{1.5cm}   F_2=\eta_V=\frac{1}{2}(n_s-1+\frac{3r}{8})=-0.014295.
\end{equation}
Considering the refined swampland conjecture \eqref{31}, we find
\begin{equation}\label{143}
C_{1}\leq 0.044094,  \hspace{1.5cm}   C_2 \leq 0.014295,
\end{equation}
$C_1$ and $C_2$ are not constants of the order of $\mathcal{O}(1)$, so they contradict the refined swampland conjecture. We now examine model II by the further refining swampland conjecture. By placing the equation \eqref{142} into \eqref{135}, we obtained
\begin{equation}\label{144}
(0.044094)^q+ 0.014295 a \geq 1-a \hspace{1cm} or \hspace{1cm} (0.00436)^q \geq 1- 1.014295 a.
\end{equation}
We can find $a$ to satisfy the recent further conjecture
\begin{equation}\label{145}
\frac{1}{1.014295}[1-(0.044094)^q]\leq a <1 ,  \hspace{1cm} q>2.
\end{equation}
Like the previous model, further refining swampland conjecture can be satisfying met for model II.
Equation \eqref{144} holds for all values of $q > 2$ with respect to $a < \frac{1}{1.014295}= 0.985906 $.
We can give an example for $a,b,q$ variables that met for this model. By selecting $q=2.1$, and using relations \eqref{144} and \eqref{145} we have $0.984503 \leq a < 0.985906 $. By choosing the $a=0.985013$, we will have $b=1-a=0.014987$.
\subsubsection{Model III}
We follow the similar route, so with respect to $n_{s}=0.964569$, $r=0.06836$ and by placing these values in equation \eqref{137}, we find
\begin{equation}\label{146}
F_1=\sqrt{2\epsilon_V}=\sqrt{\frac{r}{8}}=0.092439,  \hspace{1.5cm}   F_2=\eta_V=\frac{1}{2}(n_s-1+\frac{3r}{8})=-0.004898.
\end{equation}
So, we obtain
\begin{equation}\label{147}
C_{1}\leq 0.092439,  \hspace{1.5cm}   C_2 \leq 0.004898,
\end{equation}
where $C_1=C_2\neq \mathcal{O}(1)$, so they are in contradiction with the refined swampland conjecture.
We now examine the model by the further refining swampland conjecture. By placing the equation \eqref{146} into \eqref{135}, we obtained
\begin{equation}\label{148}
(0.092439)^q+ 0.004898 a \geq 1-a \hspace{1.5cm} or \hspace{1.5cm} (0.092439)^q \geq 1- 1.004898 a.
\end{equation}
To satisfy the conjecture we obtain the $a$
\begin{equation}\label{149}
\frac{1}{1.004898}(1-(0.092439)^q)\leq a <1 ,  \hspace{1cm} q>2,
\end{equation}
here, further refining swampland conjecture can be satisfying.
Equation \eqref{148} holds for all values of $q > 2$ with respect to $a < \frac{1}{1.01686}= 0.995126 $.
We can give an example for the $a,b,q$ variables.
By selecting $q=2.1$, and using relations \eqref{148} and \eqref{149} we have $0.988424 \leq a < 0.995126 $. Also by obtaining the $a=0.992156$, we will have the $b=1-a=0.007844$.

\section{Stability}
Here, we check the stability of the model for one case of mimetic gravity, namely, model III. However, following the same analysis, one can check the stability of the rest two cases as well.
In general, there are different ways to study the stability of a model. In cosmology, the stability of different methods is examined, e.g., Refs.
 \cite{17,85,86}. However, we use the speed of sound method to evaluate the stability of our considered model.
Generally, stability studied by the method of sound speed  is described by
\begin{equation}\label{35}
C_{s}^{2}=\frac{\textrm{d}P}{\textrm{d}\rho}>0.
\end{equation}It means the stability conditions occur when $C_{s}^{2}$ is positive. To plot the diagram, we need to obtain the energy density and pressure so that we will have:
\begin{equation}\label{36}
\begin{split}
\rho =&\frac{\bigg(X+Y\Big((X+Y)^{1-d}+Yt(1-d)\Big)^{\frac{d}{1-d}}\bigg )^{2d} }{1+4\delta+3\delta^{2}}\bigg(3+ 3\delta-\frac{Yd\delta  \Big((X+Y)^{1-d}-Yt(-1+d)\Big)^{\frac{1-d}{d}}}{(X+Y) } \bigg),
\end{split}
\end{equation}
and
\begin{equation}\label{37}
\begin{split}
 P=&-\frac{1}{-\delta^{2}+(1+2\delta)^{2}}\bigg[(3+3\delta)\bigg(X+Y\Big(\big((X+Y)^{1-d}-Yt(-1+d)\big)^{\frac{1}{1-d}}\Big)^{d}\bigg)^{2d}
\\
&+Y(2+3\delta)\bigg(X+Y\Big(\big((X+Y)^{1-d}-Yt(-1+d)\big)^{\frac{1}{1-d}}\Big)^{d}\bigg)^{-1+2d}d\bigg].
\end{split}
\end{equation}

\begin{figure}[h!]
 \begin{center}
 \includegraphics[height=6cm,width=6cm]{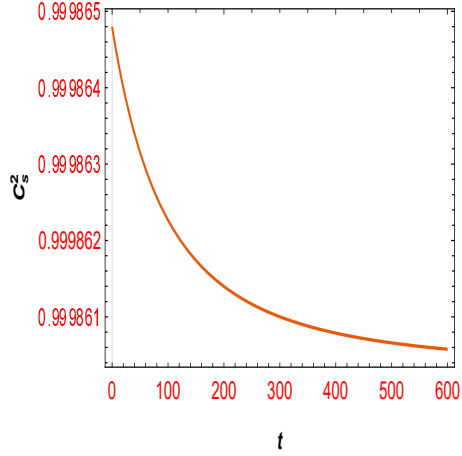}
 \caption{square of sound speed versus cosmic
time.}
 \label{fig5}
 \end{center}
 \end{figure}
\hspace{-0.6cm}As shown in figure  \ref{5}, the stability of our model is evident throughout the evolution of the universe, and this stability is apparent at all times. In the stability figures, the unit of cosmic time is usually referred to as billion years. In fact, if we consider the cosmic time coordinates as billion years, we find that our model as a whole is a stable model at all times. For a physically viable model, the sound speed should be positive and less than the speed of light $(C = 1)$. Thus the condition  $0 \leq C_{s}^{2} \leq1$ validates the acceptability of the model. Figure 5 shows that a specific constant value is obtained for large values of t

\newpage
\section{Conclusion}
The mimetic $f(R, T)$ gravity model is examined from the swampland conjecture point of view. For this purpose, we introduced several inflation solutions of the Hubble parameter such  as $(XN^{a}+YN\exp(b N))^{d}$, $(Xc^{N}+YN^{a})^{d}$ and $(X+YN)^{d}$ in  $f = R+\delta T$ gravity model.   Then, we calculated quantities such as potential and  Lagrange multiplier, slow-roll, and other cosmological parameters. We also studied the scalar spectrum index and the tensor to the scalar ratio for each model separately.
Also, by adjusting the free parameters of each model, acceptable values for these two crucial cosmological parameters, namely $n_{s}$ and $r$, were determined, which also corresponded to the latest observable data such as Planck data and BICEP2/Keck Array data.
Then, we selected the third model to examine the compatibility of the mimetic f (R, T) inflationary model with the recently introduced refined swampland dS conjecture.
We calculated the refined swampland dS conjecture concerning the potential and its first and second-order derivatives. Since the study of each of the first and second components of the swampland conjecture viz $C_{1}$ and $C_{2}$ in terms of two important cosmological parameters, namely $n_{s}$ and $r$, can be significant, so according to the mentioned equations, we created structures such as $C_{1, 2}-n_{s}$ and $C_{1, 2}-r$. By plotting some figures, we determined the allowable range related to each of these cosmological parameters and the components of the swampland conjecture.
We also studied a new constraint that is formed according to the combination of two components of the swampland as a function $f \geq C_{1}^{2}C_{2}^{2}$ 
and we determined the allowable range of the swampland component that specifies the $C_i$ that can be acceptable for the model. Thus the virtual range is specified as $C_{i} \leq 1.26911 \times 10^{-6}$.
The allowable range for these calculation of model III was in a small window in figure 4.  Then, we challenged the future refining swampland conjecture with these models. We determined all models contradict with swampland ds conjuncture since their coefficients are not unite order,i.e., $C_{1}=C_{2}\neq \mathcal{O}(1)$. Then by adjusting the free parameters of recent farther refining swampland conjecture viz $a, b, q $, we showed compatibility of all models with it.
In addition to the above calculations, we also had to check the stability of the model. Therefore, we studied the stability of model III for the mimetic f(R, T) gravity. We concluded that the stability of our model is evident throughout the evolution of the universe, and this stability is apparent at all times. Given the above concepts, such a construction of inflation models can be challenged with other swampland conjectures such as TCC. one can study additional limitations, which could be an idea for a more in-depth study of future work. Of course, this will be interesting to study perturbations of these models in full detail, which is the subject of future work.

\section{Appendix A: The slow-roll and spectral parameters (model I)}
By placing equation (18) into the equations (15) and (16),  we will have

\begin{equation}\label{A1}
\begin{split}
\epsilon
&=-\bigg\{\bigg(Nd(YN^{a}+Xc^{N})^{1-d}(YN^{a}+Xc^{N})^{-3d}\bigg(d(YN^{a}+Xc^{N})^{-2+2d}(YN^{-1+a}a+Xc^{N}\log(c))^{2}\\
&+(YN^{a}+Xc^{N})^{-2+d}\bigg[(-1+d)(YN^{-1+a}a+Xc^{N}\log(c))^{2}\\
&+(YN^{a}+Xc^{N})\Big(YN^{-2+a}a(-1+a)+Xc^{N}\log(c)^{2}\Big)\bigg](YN^{a}+Xc^{N})^{d}\\
&+(YN^{a}+Xc^{N})^{-1+d}(YN^{-1+a}a+Xc^{N}\log(c))(YN^{a}+Xc^{N})^{d}\bigg)^{2}\bigg)\bigg/\bigg(4(YN^{a}a+XNc^{N}\log(c))\\
&\bigg[3+d(YN^{a}+Xc^{N})^{-1+d}(YN^{-1+a}a+Xc^{N}\log(c)) (YN^{a}+Xc^{N})^{-d}\bigg]^{2}\bigg)\bigg\},
\end{split}
\end{equation}

\begin{equation}\label{A2}
\begin{split}
\eta =&-\bigg\{\bigg(-\bigg[\bigg((YN^{a}+Xc^{N})^{-4+2d}(YN^{a}+Xc^{N})^{2-2d}\bigg[(-1+d)(YN^{-1+a}a+Xc^{N}\log(c))^{2}\\
&+(YN^{a}+Xc^{N})\Big(YN^{-2+a}(-1+a)a+Xc^{N}\log(d)^{2}\Big)\bigg]^{2}\bigg)\Big/\Big(2\big(YN^{-1+a}a+Xc^{N}\log(c)\big)^{2}\Big)\bigg]\\
&+4(YN^{a}+Xc^{N})^{-2+d}(YN^{a}+Xc^{N})^{1-d}\Big/\Big(YN^{-1+a}a+Xc^{N}\log(c)\Big)\\
&\times \bigg[(-1+d)\Big(YN^{-1+a}a+Xc^{N}\log(c)\Big)^{2}+(YN^{a}+Xc^{N})\Big(YN^{-2+a}(-1+a)a+Xc^{N}\log(c)^{2}\Big)\bigg]\\
&+\frac{1}{2}d^{2}(YN^{a}+Xc^{N})^{-2+2d}\Big(YN^{-1+a}a+Xc^{N}\log(c)\Big)^{2}(YN^{a}+Xc^{N})^{-2d}+3d(YN^{a}+Xc^{N})^{-2d}\\
&\times \bigg[(-1+d)\Big(YN^{-1+a}a+Xc^{N}\log(c)\Big)^{2}+(YN^{a}+Xc^{N})\Big(YN^{-2+a}(-1+a)a+Xc^{N}\log(c)^{2}\Big)\bigg]\\
&\times (YN^{a}+Xc^{N})^{-d}+9d(YN^{a}+Xc^{N})^{-1+d}\Big(YN^{-1+a}a+Xc^{N}\log(c)\Big)(YN^{a}+Xc^{N})^{-d}\bigg)\bigg/\\
&2\bigg[3+d(YN^{a}+Xc^{N})^{-1+d}\Big(YN^{-1+a}a+Xc^{N}\log(c)\Big)(YN^{a}+Xc^{N})^{-d}\bigg]\bigg\}.
\end{split}
\end{equation}

With respect to equations (19), (20), and (17), the scalar spectral index and tensor-to-scalar ratio are calculated in the following form.
\begin{equation}\label{A3}
\begin{split}
n_{s}-1=&-2\bigg\{\bigg(-\bigg[\bigg((YN^{a}+Xc^{N})^{-4+2d}(YN^{a}+Xc^{N})^{2-2d}\bigg[(-1+d)\Big(YN^{-1+a}a+Xc^{N}\log(c)\Big)^{2}\\
&+(YN^{a}+Xc^{N})\Big(YN^{-2+a}(-1+a)a+Xc^{N}\log(c)^{2}\Big)\bigg]^{2}\bigg)\Big/\Big(2\big(YN^{-1+a}a+Xc^{N}\log(c)\big)^{2}\Big)\bigg]\\
&+4(YN^{a}+Xc^{N})^{-2+d}(YN^{a}+Xc^{N})^{1-d}\Big/\Big(YN^{-1+a}a+Xc^{N}\log(c)\Big)\\
&\times \bigg[(-1+d)\Big(YN^{-1+a}a+Xc^{N}\log(c)\Big)^{2}+(YN^{a}+Xc^{N})\Big(YN^{-2+a}(-1+a)a+Xc^{N}\log(c)^{2}\Big)\bigg]\\
&+\frac{1}{2}d^{2}(YN^{a}+Xc^{N})^{-2+2d}\Big(YN^{-1+a}a+Xc^{N}\log(c)\Big)^{2}(YN^{a}+Xc^{N})^{-2d}+3d(YN^{a}+Xc^{N})^{-2d}\\
&\times \bigg[(-1+d)\Big(YN^{-1+a}a+Xc^{N}\log(c)\Big)^{2}+(YN^{a}+Xc^{N})\Big(YN^{-2+a}(-1+a)a+Xc^{N}\log(c)^{2}\Big)\bigg]\\
&\times (YN^{a}+Xc^{N})^{-d}+9d(YN^{a}+Xc^{N})^{-1+d}\Big(YN^{-1+a}a+Xc^{N}\log(c)\Big)(YN^{a}+Xc^{N})^{-d}\bigg)\Big/\\
&\bigg[2\bigg(3+d(YN^{a}+Xc^{N})^{-1+d}\Big(YN^{-1+a}a+Xc^{N}\log(c)\Big)(YN^{a}+Xc^{N})^{-d}\bigg)\bigg]\bigg\}\\
&+6\bigg\{\bigg(Nd(YN^{a}+Xc^{N})^{1-d}(YN^{a}+Xc^{N})^{-3d}\bigg(d(YN^{a}+Xc^{N})^{-2+2d}\Big(YN^{-1+a}a+Xc^{N}\log(c)\Big)^{2} \\
&+(YN^{a}+Xc^{N})^{-2+d}\bigg[(-1+d)\Big(YN^{-1+a}a+Xc^{N}\log(c)\Big)^{2}+(YN^{a}+Xc^{N})\\
&\times \Big(YN^{-2+a}a(-1+a)+Xc^{N}\log(c)^{2}\Big)\bigg](YN^{a}+Xc^{N})^{d}+(YN^{a}+Xc^{N})^{-1+d}\\
&\times \Big(YN^{-1+a}a+Xc^{N}\log(c)\Big)(YN^{a}+Xc^{N})^{d}\bigg)^{2}\bigg)\bigg[4\Big(YN^{a}a+XNc^{N}\log(c)\Big)^{-1}/\\
&\bigg(3+d(YN^{a}+Xc^{N})^{-1+d}\Big(YN^{-1+a}a+Xc^{N}\log(c)\Big) (YN^{a}+Xc^{N})^{-d}\bigg)^{2}\bigg]\bigg\},
\end{split}
\end{equation}

\begin{equation}\label{A4}
\begin{split}
r=&-16\bigg\{\bigg(Nd(YN^{a}+Xc^{N})^{1-d}(YN^{a}+Xc^{N})^{-3d}\bigg[d(YN^{a}+Xc^{N})^{-2+2d}\Big(YN^{-1+a}a+Xc^{N}\log(c)\Big)^{2}\\
&+(YN^{a}+Xc^{N})^{-2+d}\bigg((-1+d)\Big(YN^{-1+a}a+Xc^{N}\log(c)\Big)^{2}+(YN^{a}+Xc^{N})\\
&\times\Big(YN^{-2+a}a(-1+a)+Xc^{N}\log(c)^{2}\Big)\bigg)(YN^{a}+Xc^{N})^{d}+(YN^{a}+Xc^{N})^{-1+d}\\
&\times\Big(YN^{-1+a}a+Xc^{N}\log(c)\Big)(YN^{a}+Xc^{N})^{d}\bigg]^{2}\bigg)\bigg[4\Big(YN^{a}a+XNc^{N}\log(c)\Big)^{-1}\Big/\\
&\bigg(3+d(YN^{a}+Xc^{N})^{-1+d}\Big(YN^{-1+a}a+Xc^{N}\log(c)\Big)(YN^{a}+Xc^{N})^{-d}\bigg)^{2}\bigg]\bigg\}.
\end{split}
\end{equation}

\section{Appendix B: The slow-roll and spectral parameters (model II)}
With respect to equation (21), (15) and (16), one can obtain
\begin{equation}\label{B1}
\begin{split}
\epsilon=&- d\bigg[(X^{2}N^{2a})a(-1+6N+2a d)+Y^{2}\textrm{e}^{2Nb}N^{2}\Big(-1+6N+6N^{2}b+2(1+Nb)^{2}d\Big)XY\textrm{e}^{Nb}N^{1+a}\\
&\times\bigg[N^{2}b(6+b)+a(-3+a+4d)+2N\Big(3+b+a(3-b+2b d)\Big)\bigg]\bigg]^{2} \Big/ 4N(Y\textrm{e}^{Nb}N+XN^{a})\\
&\times\bigg[XN^{a}a+Y\textrm{e}^{Nb}N(1+Nb) (XN^{a}(3N+a d))+YN\textrm{e}^{Nb}(d+N(3+b d))\bigg]^{2},
\end{split}
\end{equation}

\begin{equation}\label{B2}
\begin{split}
\eta=&-\frac{1}{4N^{2}(YN\textrm{e}^{Nb}+XN^{a})^{2}\bigg(3+\frac{a d}{N}+\frac{  (1-a+Nb)d}{ N\textrm{e}^{ XN^{a}}}\bigg)}\bigg\{18N(YN\textrm{e}^{Nb}+XN^{a})\\
&\times\Big(YNe^{Nb}(1+Nb)+XN^{a}a\Big)d+\Big(YN\textrm{e}^{Nb}(1+Nb)+XN^{a}a\Big)^{2}d^{2}\\
&+8N(YNe^{Nb}+XN^{a})\bigg(X^{2}N^{2a}a(-1+a d)+Y^{2}\textrm{e}^{2Nb}N^{2}(-1+(1+Nb)^{2}d)\bigg)\\
&+XY\textrm{e}^{Nb}N^{1+a}\Big(a^{2}+Nb(2+Nb)+a(-3+2Nb(-1+d)+2d)\Big)\bigg\}\bigg/\\
&\bigg\{XN^{a}a+YNe^{Nb}(1+Nb)+6d\bigg[X^{2}N^{2a}a(-1+a d)+Y^{2}N^{2}\textit{e}^{2Nb}(-1+(1+Nb)^{2}d)\bigg]\\
&+XY\textit{e}^{Nb}N^{1+a}\Big(a^{2}+Nb(2+Nb)+a(-3+2Nb(-1+d)+2d)\Big)-\bigg[X^{2}N^{2a}a(-1+a d)\\
&+Y^{2}N^{2}\textit{e}^{2Nb}(-1+(1+Nb)^{2}d)+XY\textrm{e}^{Nb}N^{1+a}\bigg(a^{2}+Nb(2+Nb)\\
&+a\Big(-3+2Nb(-1+d)+2d\Big)\bigg)\bigg]^{2}\Big/\Big(XN^{a}a+YN\textrm{e}^{Nb}(1+Nb)\Big)^{2}\bigg\}.
\end{split}
\end{equation}

Also, $n_{s}$ and r, with respect to the above equations, calculate in the following form

\begin{equation}\label{B3}
\begin{split}
n_{s}-1=&- \frac{1}{2N^{2}(YN\textrm{e}^{Nb}+XN^{a})^{2}\Big(3+\frac{a d}{N}+\frac{ (1-a+Nb)d}{ N\textrm{e}^{XN^{a}}}\Big)}\bigg(18N(YN\textrm{e}^{Nb}+XN^{a})\\
&\times\Big(YN\textrm{e}^{Nb}(1+Nb)+XN^{a}a\Big)d+\Big(YN\textrm{e}^{Nb}(1+Nb)+XN^{a}a\Big)^{2}d^{2}\\
&+\bigg[8N(YN\textrm{e}^{Nb}+XN^{a})\bigg(X^{2}N^{2a}a(-1+a d)+Y^{2}\textrm{e}^{2Nb}N^{2}(-1+(1+Nb)^{2}d)\\
&+XY\textrm{e}^{Nb}N^{1+a}\Big(a^{2}+NY(2+Nb)+a\big(-3+2NY(-1+d)+2d\big)\Big)\bigg)\bigg]\Big/\\
&\Big(XN^{a}a+YN\textrm{e}^{Nb}(1+Nb)\Big)+6d\bigg(X^{2}N^{2a}a(-1+a d)+Y^{2}N^{2}\textit{e}^{2Nb}(-1+(1+Nb)^{2}d)\\
&+XY\textrm{e}^{Nb}N^{1+a}\Big(a^{2}+NY(2+Nb)+a(-3+2Nb(-1+d)+2d)\Big)\bigg)-\bigg[X^{2}N^{2a}a(-1+a d)\\
&+Y^{2}N^{2}\textrm{e}^{2Nb}(-1+(1+Nb)^{2}d)+XY\textrm{e}^{Nb}N^{1+a}\bigg(a^{2}+NY(2+Nb)\\
&+a(-3+2NY(-1+d)+2d)\bigg)\bigg]^{2}\Big/\Big(XN^{a}a+YN\textrm{e}^{Nb}(1+Nb)\Big)^{2}\bigg)\\
&+6\bigg(\bigg[d\bigg[X^{2}N^{2a}a(-1+6N+2a d)\\
&+Y^{2}\textrm{e}^{2Nb}N^{2}\Big(-1+6N+6N^{2}Y+2(1+Nb)^{2}d\Big)XY\textrm{e}^{Nb}N^{1+a}\bigg(N^{2}Y(6+b)\\
&+a(-3+a+4d)+2N\Big(3+b+a(3-b+2b d)\Big)\bigg)\big]^{2}\bigg]\Big/\bigg[4N(Y\textrm{e}^{Nb}N+XN^{a})\\
&\Big(XN^{a}a+Y\textrm{e}^{Nb}N(1+Nb)\Big) \Big(XN^{a}(3N+a d)+YN\textrm{e}^{Nb}(d+N(3+b d))\Big)^{2}\bigg]\bigg),
\end{split}
\end{equation}

\begin{equation}\label{B4}
\begin{split}
r=&-4\bigg(d\bigg(X^{2}N^{2a}a(-1+6N+2a d)+Y^{2}\textrm{e}^{2Nb}N^{2}\Big(-1+6N+6N^{2}b+2(1+Nb)^{2}d\Big)XY\textrm{e}^{Nb}N^{1+a}\\
&\bigg[N^{2}b(6+b)+a(-3+a+4d)+2N\Big(3+b+a(3-b+2b d)\Big)\bigg]\bigg)^{2}\bigg)\bigg/\bigg[N(Y\textrm{e}^{Nb}N+XN^{a})\\
&\Big(XN^{a}a+Y\textrm{e}^{Nb}N(1+Nb)\Big) \Big(XN^{a}(3N+a d)+YN\textrm{e}^{Nb}\big(d+N(3+b d)\big)\Big)^{2}\bigg].
\end{split}
\end{equation}

\section{Appendix C: The swampland conjecture structure}
According to the potential of equation (32), its first and second derivatives are calculated as follows.
\begin{equation}\label{C1}
\begin{split}
V'(\phi)=&-\bigg[Y(X+Y)^{d}d\Big(\big((X+Y)^{1-d}-Y(-1+d)\phi\big)^{\frac{1}{1-d}}\Big)^{-1+2d}\bigg(Y\Big(-4+\delta\big(-5+\delta(11+6\delta)\big)\Big)\\
&(-1+2d)+18(-1+\delta)(1+\delta)(1+2\delta)\Big((X+Y)^{1-d}-Y(-1+d)\phi\Big)^{\frac{1}{1-d}}\bigg)]\bigg/\\
&\bigg[(1+\delta)(1+3\delta)\Big(X+Y-Y(X+Y)^{d}(-1+d)\phi\Big)\bigg],
\end{split}
\end{equation}
and
\begin{equation}\label{C2}
\begin{split}
V''(\phi)=&-\bigg[Y^{2}(X+Y)^{2d}d\Big(\big((X+Y)^{1-d}-Y(-1+d)\phi\big)^{\frac{1}{1-d}}\Big)^{-1+2d}\\
&\bigg(Y\Big(-4+\delta\big(-5+\delta(11+6\delta)\big)\Big)(2+d(-7+6d))+18(-1+\delta)
(1+2\delta)(1+\delta)\\
&\times(-1+3d)\Big((X+Y)^{1-d}-Y(-1+d)\phi\Big)^{\frac{1}{1-d}}\bigg)\bigg]\bigg/\\
&\bigg[(1+\delta)(1+3\delta)\Big(X+Y-Y(X+Y)^{d}(-1+d)\phi\Big)^{2}\bigg],
\end{split}
\end{equation}

\hspace{-0.6cm}by placing equations (32), (C1), and (C2) in swampland dS conjectures in equation (31), we will have.

\begin{equation}\label{C3}
\begin{split}
-&\bigg[\bigg(Y(X+Y)^{d}d\Big(\big((X+Y)^{1-d}-Y(-1+d)\phi\big)^{\frac{1}{1-d}}\Big)^{-1+2d}\bigg(Y\Big(-4+\delta\big(-5+\delta(11+6\delta)\big)\Big)\\
&(-1+2d)+18(-1+\delta)(1+\delta)(1+2\delta)\Big((X+Y)^{1-d}-Y(-1+d)\phi\Big)^{\frac{1}{1-d}}\bigg)\bigg)\Big/\\
&\bigg[(1+\delta)(1+3\delta)\Big(X+Y-Y(X+Y)^{d}(-1+d)\phi\Big)\bigg]\bigg]\Big/\\
&-\delta\rho+\frac{1}{(1+\delta)(1+3\delta)}\bigg(\Big((X+Y)^{1-d}-Y(-1+d)\phi^{\frac{1}{1-d}}\Big)^{-1+2d}\\
&\times\bigg[Y\Big(4+\delta\big(5-\delta(11+6\delta)\big)\Big)d-9(-1+\delta)(1+\delta)(1+2\delta)
\\
&\times\Big((X+Y)^{1-d}-Y(-1+d)\phi\Big)^{\frac{1}{1-d}}\bigg]\bigg)
>C_1,
\end{split}
\end{equation}

\begin{equation}\label{C4}
\begin{split}
-&\bigg[\bigg(Y^{2}(X+Y)^{2d}d\Big(\big((X+Y)^{1-d}-Y(-1+d)\phi\big)^{\frac{1}{1-d}}\Big)^{-1+2d}\\
&\bigg[Y\Big(-4+\delta\big(-5+\delta(11+6\delta)\big)\Big)\big(2+d(-7+6d)\big)+18(-1+\delta)
(1+2\delta)(1+\delta)\\
&(-1+3d)\Big((X+Y)^{1-d}-Y(-1+d)\phi\Big)^{\frac{1}{1-d}}\bigg]\bigg)\Big/\\
&\bigg((1+\delta)(1+3\delta)\Big(X+Y-Y(X+Y)^{d}(-1+d)\phi\Big)^{2}\bigg)\bigg]\bigg/\\
&-\delta\rho+\frac{1}{(1+\delta)(1+3\delta)}\bigg(\Big((X+Y)^{1-d}-Y(-1+d)\phi^{\frac{1}{1-d}}\Big)^{-1+2d}\\
&\times\bigg[Y\Big(4+\delta\big(5-\delta(11+6\delta)\big)\Big)d-9(-1+\delta)(1+\delta)(1+2\delta)
\\
&\times\Big((X+Y)^{1-d}-Y(-1+d)\phi\Big)^{\frac{1}{1-d}}\bigg]\bigg)<-C_2
\end{split}
\end{equation}
\end{document}